\documentstyle[prl,aps,twocolumn]{revtex}

\catcode`\@=11

\def\maketitle2{\par 
\begingroup
\let\cite\@bylinecite
\def\thefootnote{\fnsymbol{footnote}}%
\twocolumn[\@maketitle2\vskip2pc]%
\thispagestyle{plain}\@thanks
\endgroup
\def\thefootnote{\arabic{footnote}}%
\setcounter{footnote}{0}%
\let\maketitle2\relax \let\@maketitle2\relax
\let\@thanks\relax \let\@authoraddress\relax \let\@title\relax
\let\@date\relax \let\thanks\relax \let\@abstract\relax
\let\@pacs\relax}

\def\abstract#1{\gdef\@abstract{{\par 
\bgroup
\ifdim\prevdepth=-1000pt \prevdepth0pt\fi
\hsize\columnwidth
\dimen0=-\prevdepth \advance\dimen0 by17.5pt \nointerlineskip
\small\vrule width 0pt height\dimen0 \relax}{~~}#1\egroup}}

\def\pacs#1{\gdef\@pacs{{\par 
\bgroup
\hsize\columnwidth \parindent0pt
\ifdim\prevdepth=-1000pt \prevdepth0pt\fi
\dimen0=-\prevdepth \advance\dimen0 by20pt\nointerlineskip
\egroup} PACS numbers:~#1}}

\def\@maketitle2{
\@preprint
\@title
\ifdim\prevdepth=-1000pt \prevdepth0pt\fi
\@authoraddress
\@date
\begin{list}{}{\leftmargin=0.10753\textwidth \rightmargin=\leftmargin
\itemsep=1pc\partopsep=-1pc}
\item\@abstract
\item\@pacs
\end{list}
}

\catcode`\@=12

\begin{document}
\draft

\def\simlt{\stackrel{<}{{}_\sim}}
\def\simgt{\stackrel{>}{{}_\sim}}
\def\longbar#1{{\overline{#1}}}

\title{The Quantum Black Hole Specific Heat Is Positive}
\author
{Andrzej Z. G\'orski$^{1,2}$ and Pawel O. Mazur$^1$}
\preprint{USC-04/97}
\address{$^1$ Department of Physics and Astronomy,
University of South Carolina, Columbia, SC 29208, USA}
\address{$^2$ Institute of Nuclear Physics, Radzikowskiego 152,
31--342 Krak\'ow, Poland}
\date{\today}
\abstract
{\small{We suggest in this Letter
that the Bekenstein--Hawking black hole
entropy accounts for the degrees of freedom
which are excited at low temperatures only
and hence it leads to the negative specific heat.
Taking into account the physical degrees of freedom
which are excited at high temperatures,
the existence of which we postulate,
we compute the total specific heat of the quantum black hole
that appears to be positive.
This is done in analogy to the Planck's
treatment of the black body radiation problem.
Other thermodynamic functions are computed as well.
Our results and the success of the thermodynamic description of the
quantum black hole suggest an underlying atomic (discrete) structure of
gravitation.
The basic properties of these {\em gravitational atoms} are found.
}}
\maketitle2
\narrowtext

The notion of entropy for black holes was introduced
by Bekenstein a long time ago \cite{Bekenstein}
and it has been extensively used since then
by many authors \cite{Hawking}.
The Bekenstein entropy was defined as
\begin{equation}
S = {1\over 4} A = 4 \pi {M^2_{\cal C}} \ ,
\label{BekEntropy}
\end{equation}
where $A$ is the black hole horizon area and $M_{\cal C}$ is its total
irreducible mass--energy. This irreducible mass-energy is left invariant
by the `reversible transformations in black hole physics'
\cite{Christodoulou}. This formula also looked strangely familiar
and it was reminiscent of physical quantities like entropy
or adiabatic invariants of Hamiltonian dynamics at the same time.
This was clear to those who discovered these properties of black holes.
Here we propose to call the irreducible mass $M_{\cal C}$
and $I_{\cal C}=4{\pi}{M^2_{\cal C}}$ the Christodoulou mass and
the Christodoulou adiabatic invariant, respectively
\cite{Christodoulou}.

It has been shown that the entropy (\ref{BekEntropy})
leads to the {\em negative} specific heat for
$x={J\over E^2}<x_c=(2{\sqrt 3}-3)^{1\over 2}=0.68125004...$, where
$J$ is the hole's angular momentum and $E$ is its rest mass-energy.

 From now on we focus our attention on the classical
non-rotating Schwarzschild
black hole for which the specific heat is negative

\begin{equation}
c_{bh} = - 8 \pi E^2  < 0 \ .
\label{BekHeat}
\end{equation}
This property of negative specific heat
is frequently used to argue that the canonical ensemble fails
for gravitating
systems, as (\ref{BekHeat}) implies negative variance
$\sigma^2(\bar E)=(\longbar{\Delta E^2})$
of the normal Gaussian probability distribution
\cite{LL}.
We shall show that this problem can be avoided.

Let us emphasize the following two important physical points.
First, the applicability of thermodynamic notions, and the entropy
(\ref{BekEntropy}) in particular, suggests the underlying
well hidden atomic or discrete structure behind
the black hole dynamics.
This observation seems to be known, though not pursued consequently,
to some authors \cite{Bekenstein2,Mukhanov,Kogan,GRG19,BekMukh}.
This situation we consider analogous to that of the atomic theory
of matter at the turn of this century.
In fact, this idea has been advocated by `t Hooft \cite{GTH}
and by one of the present authors \cite{APP1,APP2}.

Second, the negative specific heat indicates that we are dealing
with an open system and some of the physical degrees
of freedom have been neglected.
These are the basic physical observations behind this
paper. Our aim is to find thermodynamic functions
of the quantum black hole, and in particular,
its total specific heat $c_{tot} > 0$, entropy,
and the partition function $Z$.
For simplicity we consider
a non--rotating black hole only because it is quite straightforward
to generalize our results to other more general black holes.
We claim that the Bekenstein--Hawking entropy
takes into account the low
temperature physical degrees of freedom whereas
the high temperature
degrees of freedom are neglected. This results
in the physically
unacceptable negative total specific heat (\ref{BekHeat}).

In fact, this situation reminds us of the Wien
law for the black body
radiation. This prompted us to ask the question:
What does the high temperature regime
of quantum black hole behavior look like?
Clearly, our understanding of this regime was missing
(an analog of the Rayleigh--Jeans regime of the black body
radiation) before we had realized that the answer is already
there, but it was somehow well hidden until only recently \cite{APP2}.
The sole fact that the concept of entropy was invoked
in the context of black hole physics should be interpreted
as saying that
we should take the fundamental idea
of {\em Atomic Hypothesis} \cite{APP1,APP2} seriously.

Hence, our idea is:
{\it (i)}   to take into account the high
temperature degrees of freedom,
{\it (ii)}  to find out a counterpart of
the Planck's interpolation
formula \cite{Planck}, and
{\it (iii)} to compute the complete
thermodynamic functions of
the quantum black hole. Finally, we shall show that
our results imply underlying
discrete or atomic structure behind
a gravitating object (the quantum black hole).
To this end, we shall follow the method
of Planck and Einstein
\cite{Planck,fluctuat}, and we shall use
a simple dimensional arguments.
In view of what was said above, the total specific heat
can be expressed as a sum of the Bekenstein--Hawking specific
heat (\ref{BekHeat}) and the specific heat
coming from the high temperature degrees of freedom,
and it should be positive:
$ c_{tot} = c_{bh} + c_{nc} > 0 $,
where the subscript $nc$ stands for {\em non-collapse}.
The meaning of the {\em non-collapse hypothesis} will be
explained in the final section of this Letter.
We expect that the specific heat of
the high temperature degrees of
freedom, $c_{nc}=c_{nc}(T)$,
is a slowly varying function of $T$. This is not unlike the
elementary example of the {\em `Einstein crystal'}, by which we mean
the single frequency crystal \cite{LL}.

We start with the Einstein
fluctuation--dissipation theorem \cite{fluctuat}
for energy fluctuations
\begin{equation}
\sigma^2_{tot}(E) =  \longbar{\Delta E^2_{tot}} =
\left( - {\partial^2S_{tot}\over
\partial E^2} \right)^{-1} = \ T^2 c_{tot}
\label{TotalFluct}
\end{equation}
that is valid for any closed quantum
atomistic system and leads to the
Gaussian probability distribution
\begin{equation}
P(\Delta E) \sim \exp
\left( - (\Delta E)^2 / 2\sigma^2
\right) \ .
\label{GaussProb}
\end{equation}
Treating the low temperature energy fluctuations
as one random Gaussian variable and
the high temperature fluctuations as the second independent
Gaussian variable we see that the total distribution must be Gaussian.
In fact, it is the convolution of both distributions
\begin{eqnarray}
P_{tot}(\Delta E) = &&\int_{-\infty}^{+\infty} dy \
P_{bh}(\Delta E - y)  \ P_{nc}(y) \sim \nonumber \\
&&\sim \exp \left( - (\Delta E)^2 /2\sigma^2_{tot} \right)
\label{convolution}
\end{eqnarray}
and, as a consequence, we have
\begin{equation}
\sigma_{tot}^2 = \sigma^2_{bh} + \sigma^2_{nc}
\label{sigmatot} \ ,
\end{equation}
where for the Bekenstein--Hawking fluctuations we have formally
\begin{equation}
\sigma_{bh}^2 = - {b \mu^2 \over 8\pi} \ ,
\label{sigmabh}
\end{equation}
where $\mu$ is the Planck mass, and $b$ is a dimensionless constant
\cite{units}.

Now, the fundamental question arises:
what expression should we choose for
the high energy, or rather, for the high temperature
fluctuations $\sigma^2_{nc}$?
In fact, we have two constants
of the dimension of energy:
the Planck energy $\mu$, that
has been already used in (\ref{sigmabh}),
and the total energy $E$ of the black hole.
No other dimensional constants
can be ascribed to a non--rotating gravitating object
(Schwarzschild black hole). This was also the reason why
it was considered earlier by one of us
\cite{APP2,talks}.
In order to understand the motivation beyond this choice we have
to recall \cite{Sommerfeldbook} that with the introduction by Planck
of his constant $h$ all fundamental constants were put in place and we
have received from Planck the system of natural units.
What is then left for us if not a large integer $N$?
The meaning of that large integer
$N$ will become clear later on \cite{GRG19,APP2,TBP,talks}.

Hence, we can assume the following
hypothesis \cite{remark1}
\begin{equation}
\sigma_{nc}^2 = a E^2  \ ,
\label{sigmaHT}
\end{equation}
where $a$ is a {\em dimensionless}
constant and $E$ is the black hole
mass--energy.
The numerical value of the constant
$a$ and its relation to an integer $N$ mentioned above
will be given later, together
with the discussion of our result within
the context of physical ideas presented first in
\cite{APP1,APP2,talks}.
Our complete formula for the energy
fluctuations now reads
\begin{equation}
\sigma_{tot}^2 = -{b\mu^2\over 8\pi}
+ a E^2  \equiv
a ( E^2 - E_0^2)   \ ,
\label{sigmatotal}
\end{equation}
where the constant $E_0$ is defined as
$E_0^2 = {b \mu^2 \over 8\pi a} \ .$
The equation (\ref{sigmatotal}) is
the basic formula we shall use to
derive other thermodynamic functions.

{\em Thermodynamic functions of the quantum black hole.}
It is a matter of simple integration to obtain from equations
(\ref{TotalFluct}) and (\ref{sigmatotal}),
after imposing the proper boundary condition,
$T \to \infty$ when $E \to \infty$:
\begin{equation}
{\partial S\over \partial E} = \beta \equiv {1\over T} =
{1\over 2aE_0} \ln {E+E_0\over E-E_0} \ .
\label{beta}
\end{equation}
This implies the temperature dependence of the mean energy
\begin{equation}
E(\beta) = E_0 \coth (aE_0\beta )  \ ,
\label{EofT}
\end{equation}
which we immediately recognize as the expression known from textbooks
\cite{LL}. The low temperature asymptotics, $T \rightarrow 0$,
\begin{equation}
E \sim E_0 + 2E_0 \ e^{-2a\beta E_0} \rightarrow E_0  \ .
\label{EofTzero}
\end{equation}
also explains the meaning of $E_0$.
Thus, the constant $E_0$ should be interpreted
as the minimal energy of the
whole system, {\em i.e.} the total
{\em zero point energy} of the quantum black hole.
The total energy $E(T)$ is bounded from below by $E_0$,
$E(T) \ge E_0$.
For the high temperature regime, as for the Planck distribution,
we have the following asymptotics, $T \rightarrow \infty$,
\begin{equation}
E \sim {1\over a} \  T \ + \ {1\over 3} a E_0^2 \ {1\over T}
\rightarrow \infty \ .
\label{EofTinfty}
\end{equation}
The leading term in this formula is clearly the Rayleigh-Jeans
result and it shall be interpreted as the energy equipartition rule.
Now, using the standard formula
for the canonical ensemble:
$E = -\partial/\partial\beta \ln Z[\beta]$
one has the partition function in the following simple form
\begin{equation}
Z[\beta] = \left( {1\over 2\sinh {\beta\epsilon\over 2} }
\right)^{1/a}   \ ,
\label{partitionfun}
\end{equation}
where $\epsilon = 2aE_0$ has been defined as
\begin{equation}
\epsilon = \mu \ \sqrt{ab\over 2\pi}  \ .
\label{epsilondef}
\end{equation}

Integrating (\ref{beta}) once again we obtain
the total entropy as a function of energy
\begin{equation}
S(E) = S_0 + {1\over 2a} {E\over E_0} \ln{E+E_0\over E-E_0} +
 {1\over 2a} \ln{E^2-E_0^2\over E_0^2}  \ ,
\label{EntropyOfE}
\end{equation}
where $S_0$ is an integration constant independent of $E$ and $a$.
It is convenient to set it to zero. This point will be discussed
shortly later.
Eq. (\ref{EntropyOfE}) gives the following asymptotics
\begin{eqnarray}
&&S(E) \sim {1\over a} \ \ln{E\over E_0} \rightarrow \infty \quad
(E \rightarrow \infty) \ ,
\nonumber \\
&&S(E) \rightarrow {1\over a} \ln 2 \quad (E \rightarrow E_0)
\ .
\label{EntropyElimits}
\end{eqnarray}
The entropy $S$ can be computed as a function
of temperature $T$, $\beta=1/T$, directly from
the partition function (\ref{partitionfun}).
Up to a constant $S_0$ mentioned earlier this gives
\begin{equation}
S(\beta) = -{1\over a} \ \ln \sinh {\beta\epsilon\over2} +
{\beta\epsilon\over2a} \
\coth{\beta\epsilon\over2} \ .
\label{EntropyOfBeta}
\end{equation}

Finally, from (\ref{beta})
we calculate the total specific heat
\begin{equation}
c_{tot} =   {a\ E_0^2 \ \beta^2 \over \sinh^2 aE_0 \beta} \ .
\label{SpecHeat}
\end{equation}
$c_{tot}$ is always positive and has
the following asymptotics
\begin{eqnarray}
&&c_{tot}(T) \sim 4 a E_0^2 \ {1\over T^2} \
e^{-2aE_0/T} \rightarrow 0 \quad
(T \rightarrow 0) \ ,
\nonumber \\
&&c_{tot}(T) \rightarrow {1\over a} - {1\over 3} {a\ E_0^2 \over T^2}
\quad (T \rightarrow \infty) \ . \label{HeatAsymp}
\end{eqnarray}

{\em The gravitational quanta.}
In order to discuss the physical consequences of the results
from the previous section,
we have to determine first what is the physical meaning of
the dimensionless constant $a$. Our choice is
\begin{equation}
a  = {1\over N} \ , \qquad N - \hbox{\rm integer} \ .
\label{adef}
\end{equation}
There are several arguments in favor of this choice.
From the point of view of this Letter
the most important fact is that the statistical
sum (\ref{partitionfun}) can be
factorized as a product of $N$
independent one--particle partition functions $Z_1$.
Now, we can rewrite (\ref{partitionfun}) as
\begin{equation}
Z  = Z_1^N \ , \qquad Z_1 \equiv  {1\over 2\sinh
{\beta\epsilon\over 2} }    \ .
\label{ZandZ1}
\end{equation}
The choice of the integer value for $N$ is forced on us
by the probabilistic interpretation. An integer $N$ is also quite
appealing on aesthetical grounds. Whenever an integer appears
naturally from some simple hypothesis we should always be on the lookout
for something fundamental \cite{Planck,Sommerfeldbook,fluctuat}.
Finally, eq. (\ref{adef}) is in agreement with the
{\em new gravitational noncommutative mechanics} introduced
in \cite{APP1,APP2,talks,TBP}, where
the wider spectrum of arguments has been given. The basic ideas
of the {\em new gravitational mechanics} \cite{APP1}
were discussed also in the context
of {\em gedanken experiments} \cite{CashNuss} before
\cite{APP1} was published. The {\em non-collapse hypothesis}, which was
implemented by {\em the postulate of the noncommutative Manin torus
replacing the uniformizing complex torus in the GRT Kepler problem
upon transition to the new gravitational mechanics} \cite{APP1}, was
tested with the thought experiments using the Planckian energy
gedanken accelerators \cite{CashNuss}.

The constants
$E_0$
(the total energy of the black
hole in the zero temperature limit) and $\epsilon$ can be expressed as
\begin{eqnarray}
E_0 \ &&= \ N \ {\epsilon\over 2} \ = \
\sqrt{N} \ \mu \ \sqrt{b \over 8 \pi}  \ ,
\label{T0energy}  \\
\epsilon \ &&= \ {1\over \sqrt{N}} \ \mu \
\sqrt{b \over 2 \pi} \ .
\label{epsilonN}
\end{eqnarray}
One can see immediately from (\ref{ZandZ1}) that $Z_1$ is
the partition function of a quantum harmonic oscillator
\begin{equation}
Z_1 \ \equiv \ {e^{-\beta\epsilon/2} \over 1 - e^{-\beta\epsilon}} \ =
\
\sum_{n=0}^{\infty}  \ e^{-\beta E_n} \ ,
\label{oscilZ1}
\end{equation}
with the energy levels $E_n$ defined as
\begin{equation}
E_n \ = \ \left( n + {1\over2} \right) \ \epsilon \ .
\label{En}
\end{equation}
Hence, our gravitational quanta are bosons: they do obey
the Bose--Einstein statistics
and eq. (\ref{EofTinfty}) states that at high temperatures
we have the equipartition of energy in the leading order. Obviously,
$\epsilon$ is the harmonic oscillator level spacing which depends
universally on $N$ \cite{APP2}.

At this point we can go back to the formula (\ref{EntropyOfE}) for
the entropy and,
in particular, to its zero temperature limit (\ref{EntropyElimits}).
We can see now that the choice of the constant
of integration, $S_0 = 0$, was quite fortunate. Indeed,
the remaining term with the overall factor of $N$
\begin{equation}
S(0) \ = \ N  \ln 2 \ ,
\label{Sof0}
\end{equation}
can be easily connected to the ground state
degeneracy $d_N = e^{S(0)}$.
From (\ref{Sof0}) we have $d_N = 2^N$. This is quite a remarkable
formula as it implies that each elementary gravitational quantum
can be in {\em two} fundamental states \cite{dN}.
We simply encounter here a $Z_2$ quantum number. An analogy to the spin
variables invites itself quite naturally. However, we {\em do not}
suggest here at this point that this double--valuedness is related
to the usual notion of spin of `elementary particles'
as our {\em gravitational atoms} are much more fundamental than the
so-called `elementary particles' and such a proposal would be
presumably too far fetched. The {\em gravitational atoms} seem to be
a part of the physical objective reality \cite{APP2}
which should find its confirmation in observations
and experimental data. One way of finding out how real they are is to
propose an experiment in which
their existence will be tested indirectly,
in the same way as the theory of Brownian motion \cite{LL} due to
Einstein, and Smoluchowski, has led to the experimental confirmation of
the existence of atoms and molecules. The {\em gravitational atoms}
do exist on a deeper level of the physical
reality. The natural scale for gravitational atoms
is the Planck scale.

We can recognize now, that the formula for $E_0$ and (\ref{adef})
imply the mass--energy quantization of the type derived first
in \cite{GRG19}:
$E^2(0) = M_0^2 = P_\mu P^\mu = m^2 N$, with some Planckian mass scale
$m$. These considerations suggest to us that the invariant mass
$M_0^2 = P_\mu P^\mu$
is quantized and it can be viewed as consisting of $N$ Planckian
mass scale  {\em gravitational atoms} (see eq. (\ref{T0energy})):
$ M_0^2 \ = \ N \ m^2 $,
where $m^2 = {b \mu^2 \over 8 \pi}$.

One can interpret this last formula as saying that at the zero
temperature the quantum system which we call here
the quantum black hole
behaves as $N$ free identical gravitational atoms of mass $m$.
We can see that it is the invariant mass squared which is
additive at the zero temperature.
It is perhaps better to say that {\em gravitational quanta} are
collective excitations in the system of $N$ {\em gravitational atoms}.
From (\ref{epsilonN}) and (\ref{oscilZ1}) it is also clear
that the large $N$ limit ($N \rightarrow \infty$)
or, equivalently, $1/N \rightarrow 0$
corresponds to the classical limit.
In this limit we get a massive gravitating object and
Einstein's general relativity is recovered. This should be
understood properly, as the {\em classical limit} ${{G\over
c^2}=K\rightarrow 0}$ is highly nontrivial.
This is similar to the transition
from quantum mechanics to classical mechanics.

Finally, we would like to comment on the {\em non--collapse hypothesis}
which was mentioned above. It is clear from the derivation
of the fundamental results in this Letter that the negative specific
heat and the gravitational collapse are intimately connected
\cite{APP2,TBP}.
Also, the postulate of the missing degrees of freedom
presented in this Letter implies directly that
the unitarity requirement
is broken in any quantum theory which does not take them into account.
Hence the $S${\em--matrix} postulate of `t Hooft \cite{TH}
is similar if not equivalent to the {\em non--collapse hypothesis}
first proposed in \cite{APP1,APP2,talks,TBP}.
The {\em non--collapse hypothesis} implies
that elastic channels for collision of `small black holes' are present
and therefore something akin to equipartition of energy is
possible in some range of temperatures.
The exact relationship between the {\em non--collapse hypothesis},
which is presented in this Letter in the context of the
fluctuation--dissipation theorem, and the issue of unitarity is quite
subtle and it is clearly beyond the scope of this Letter. However, in
view of the reaction this point has received
recently \cite{TBP,talks}, we plan the more detailed pedagogical
paper in the nearest future.


\vskip3.0pt
POM work was supported by the NSF grant to the University of South
Carolina. AZG thanks the Kosciuszko Foudation for the Fellowship
and the Department of Physics and Astronomy,
University of South Carolina for its hospitality.


\vspace{-0.5cm}
\begin{references}
\vspace{-1.4cm}





\bibitem{Bekenstein} J. D. Bekenstein, {\em Phys. Rev.}
{\bf D7} 2333 (1973).

\bibitem{Hawking} S. Hawking {\em Nature} {\bf 248} 30 (1974);
{\em Comm. Math. Phys.} {\bf 43} 199 (1975).

\bibitem{Christodoulou} D. Christodoulou, {\em Phys. Rev. Lett.}
{\bf 25} 1596 (1970).

\bibitem{LL} L.D. Landau, E.M. Lifshitz, {\em Statistical
Physics, A Course in Theoretical Physics Vol. V}
(Nauka, Moscow 1964).

\bibitem{Bekenstein2} J.D. Bekenstein, {\em Lett. Nuovo Cim.} {\bf 11}
467 (1974).

\bibitem{Mukhanov} V. Mukhanov, {\em JETP Letters}
{\bf 44} 63 (1986).

\bibitem{Kogan} Ya. I. Kogan, {\em JETP Letters} {\bf 44} 267 (1986);
preprint OUTP-94-39-P, Dec. 94, 13 pp., hep-th/9412232.

\bibitem{GRG19} P.O. Mazur, {\em GRG} {\bf 19} 1173 (1988).

\bibitem{BekMukh} J.D. Bekenstein, V. Mukhanov, {\em Phys. Lett.}
{\bf B360} 7 (1995).

\bibitem{GTH} G. 't Hooft, {\em Quantum mechanical behavior
in a deterministic model}, quant--ph/9612018.

\bibitem{APP1} P.O. Mazur, {\em Acta Phys. Pol.} {\bf B26}
1685 (1995).

\bibitem{APP2} P.O. Mazur, {\em Acta Phys. Pol.} {\bf B27}
1849 (1996).

\bibitem{Planck} M. Jammer, {\em The Conceptual Development of Quantum
Mechanics} (McGraw-Hill NY 1966), Chapter 1.

\bibitem{fluctuat} A. Einstein, {\em Ann. Phys.} {\bf 20}
199 (1906); {\em Phys. Z.} {\bf 10} 185, 817 (1909).

\bibitem{units} Throughout our Letter we use natural units with
$\hbar = c = G = k = 1$. Hence, the Planck mass $\mu = 1$. However,
we write down $\mu$ explicitly to keep the dimensionality
transparent.

\bibitem{TBP} P.O. Mazur, talk at the Conference BSM5, Bergen (April
1997), to be published.

\bibitem{talks} P.O. Mazur, lecture at the Summer Institute,
USC (Summer 1995),
seminars at Caltech, UCLA (February
1996) and Tel--Aviv University (March 1996).

\bibitem{Sommerfeldbook} A. Sommerfeld, {\em Thermodynamics and
Statistical Mechanics,
Lectures on Theoretical Physics Vol. V} (Academic Press 1988).

\bibitem{remark1} This hypothesis is very much alike the Planck's
hypothesis for the black body radiation. He had fluctuations of the
form: $\alpha E^2 + \beta E$. This is equivalent to
$\alpha {E^\prime}^2 + \gamma$ with $\gamma = -\beta^2/4\alpha$
and the energy shifted by: $E^\prime = E + \beta/2\alpha$.
Actually, both functions are quadratic in energy.

\bibitem{dN} The total number of states for a system with $N$ constituents
can be written as: $d_N = \sum_{k=0}^\infty \left( {N\atop k} \right)
\equiv
\sum^\infty_{k=0} \left( {N \atop k} \right) 1^k 1^{N-k} \equiv
(1+1)^N \equiv 2^N$.

\bibitem{TH} G. `t Hooft, {\em Unitarity of the Black Hole Scattering
Matrix}, {\em in Quantum Coherence and Reality}, Ed. J. S. Anandan and
J. L. Safko, {\em World Scientific, Singapore, 1994}.

\bibitem{CashNuss} A. Casher and S. Nussinov, {\em Some Speculations on
the Ultimate Planck Energy Accelerators}, preprint IASSN-HEP-95/84,
October 1995, hep-ph/9510364.






\end{references}
\end{document}